# Knowledge Graph Based Waveform Recommendation: A New Communication Waveform Design Paradigm

Wei Huang, Tianfu Qi, Yundi Guan, Qihang Peng, Jun Wang
National Key Laboratory of Science and Technology on Communications,
University of Electronic Science and Technology of China, Chengdu, China

*Abstract*—**Traditionally, a communication waveform is designed by experts based on communication theory and their experiences on a case-by-case basis, which is usually laborious and time-consuming. In this paper, we investigate the waveform design from a novel perspective and propose a new waveform design paradigm with the knowledge graph (KG)-based intelligent recommendation system. The proposed paradigm aims to improve the design efficiency by structural characterization and representations of existing waveforms and intelligently utilizing the knowledge learned from them. To achieve this goal, we first build a communication waveform knowledge graph (CWKG) with a first-order neighbor node, for which both structured semantic knowledge and numerical parameters of a waveform are integrated by representation learning. Based on the developed CWKG, we further propose an intelligent communication waveform recommendation system (CWRS) to generate waveform candidates. In the CWRS, an improved involution1D operator, which is channel-agnostic and space-specific, is introduced according to the characteristics of KG-based waveform representation for feature extraction, and the multi-head self-attention is adopted to weigh the influence of various components for feature fusion. Meanwhile, multilayer perceptron-based collaborative filtering is used to evaluate the matching degree between the requirement and the waveform candidate. Simulation results show that the proposed CWKG-based CWRS can automatically recommend waveform candidates with high reliability.**

## I. Introduction

For wireless communications, a communication waveform is usually composed of multiple access and duplex scheme, frame/slot structure, modulation, channel coding, and interleaving scheme, etc [1]. The task of communication waveform design is to determine these schemes and corresponding parameters so that the information transmission requirement under specific application scenarios and resource constraints can be efficiently met. For a new requirement, the communication waveform is commonly designed by experts with solid communication theory knowledge and rich experiences, which is usually laborious and time-consuming. For example, it takes thousands of experts worldwide several years to design 4G and 5G waveforms. Moreover, the optimality of the proposed waveform could not always be guaranteed due to theoretical and empirical limitations. Therefore, it is well-motivated to improve the efficiency of waveform design.

Model-based optimization approaches have been widely used to facilitate the design of communication waveforms. The adaptive modulation and coding (AMC) [1] is a typical one. In the regime of software-defined radio (SDR) and cognitive radio (CR), a spectrally modulated, spectrally encoded (SMSE) framework [2] is proposed to represent and analyze orthogonal frequency division multiplexing (OFDM)-based communication waveform. In [3], a framework for joint dynamic resource allocation and waveform adaptation is presented based on generalized signal expansion functions. With the development of machine learning, data-driven methods are proposed, for which the transceiver is considered a black box and part or all of the components are replaced by deep neural networks (DNNs) [4], [5]. Recently, the combination of model-driven and data-driven methods has been proposed for better performance, e.g., ComNet, RTN, ViterbiNet [6], [7].

However, for the above methods, the communication waveform is not considered as a whole to optimize. Most of the existing works focus on the design of some components, e.g., the modulation and channel coding, while other crucial parts are not jointly optimized, e.g., the frame/slot structure. Moreover, the knowledge of the existing communication waveforms already successfully applied in different scenarios can be leveraged when designing new waveforms in similar scenarios.

Recently, knowledge graph-based recommendation systems (KGRS) have attracted considerable research and development efforts and have been successfully applied in the fields of search, entertainment, and shopping [8], [9] due to the ability to improve recommendation accuracy by utilizing structured information. Inspired by the KGRS, we propose a communication waveform knowledge graph (CWKG) aided communication waveform recommendation system (CWRS).

In this paper, a new waveform design paradigm based on the CWKG is proposed, which facilitates the efficiency of communication waveform design and is embedded with the capability of utilizing the knowledge from existing waveforms. A knowledge graph is a structured representation of facts, consisting of entities, relationships, and semantic descriptions [8]. The existing waveform knowledge is characterized and represented by the knowledge graph, and the CWKG is built by incorporating both the waveform scheme and the corresponding performance under a specific environment. Based on the CWKG, we further propose an intelligent CWRS to generate waveform candidates according to the information transmission requirement, such as the application environment and quality-of-service (QoS). We adopt an improved involution1D operator and multi-head self-attention to perform embedding representation enhancement (ERE) for feature extraction and fusion. Meanwhile, we use the multilayer perceptron (MLP)-based collaborative filtering (CF) to evaluate the degree how the waveform candidate matches with the requirement. Simulation results show that the proposed CWKG based CWRS can recommend waveform candidates with high reliability.

The remainder of this paper is organized as follows. First, the framework of CWKG based CWRS is presented in Section II. Then, details of CWKG and CWRS are given in Section III and IV, respectively. Next, we show simulation results in Section V. Finally, we conclude in Section VI and discuss future research directions.

## II. The Framework of KG-Based Communication Waveform Recommendation System

The framework of CWKG aided CWRS is shown in Fig. 1. A CWKG is built with structured representation of existing communication waveforms, so as to systematically utilize the waveform knowledge. According to the information transmission requirement, the CWRS generates waveform candidates based on the CWKG. The performance of each waveform candidate is evaluated through the evaluation system. If one waveform candidate meets the requirement, it will be accepted and added to the CWKG. If the obtained waveform candidates could not fully meet the requirement under some particular environment, they can serve as foundations for further optimization to accelerate the design process, in which the existing model-based and data-based methods can be applied. As CWRS is the core of this new waveform design paradigm, this paper focuses on the CWKG and CWRS of this framework.

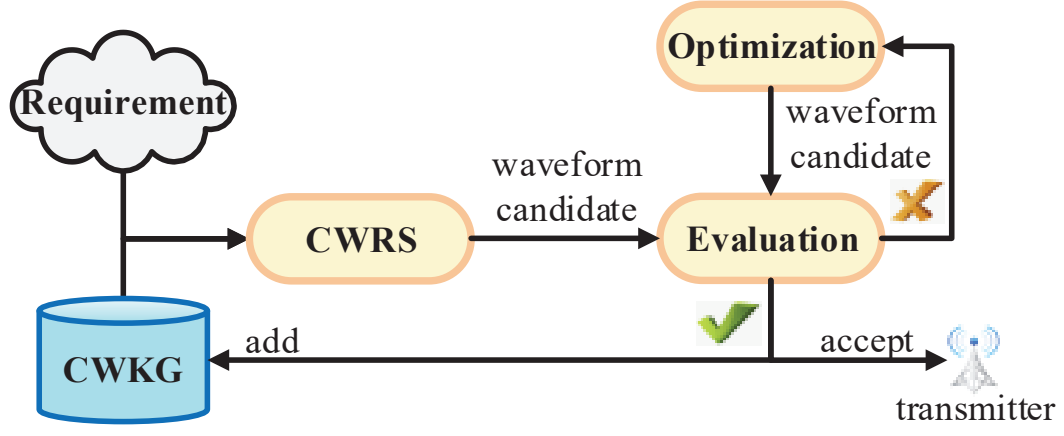

Fig. 1. The framework of CWKG aided CWRS

Mathematically, the waveform recommendation task is defined as "***Given an environment*** $u$ ***and a target waveform*** $v$***, the CWRS estimates the probability score of the matching degree between them***." That is,

$$\hat{y}_{uv} = f_{\Theta}(u, v) \tag{1}$$

where $f_{\Theta}$ represents the neural network with parameter $\Theta$ adopted in the CWRS, and $\hat{y}_{uv}$ denotes the probability score measuring the availability of target waveform $v$ under environment $u$.

As illustrated in Fig. 2, the framework of the CWRS can be divided into three parts: Knowledge Representation Learning (KRL), Embedding Representation Enhancement (ERE), and Collaborative Filtering (CF). Correspondingly, $f_{\Theta}(\cdot)$ can be divided into $f^1_{\Theta_1}(\cdot), f^2_{\Theta_2}(\cdot), f^3_{\Theta_3}(\cdot)$ as following:

$$\begin{aligned} (\mathbf{E}_u, \mathbf{E}_v) &= f^1_{\Theta_1}(u, v) \\ (\mathbf{Z}_u, \mathbf{Z}_v) &= f^2_{\Theta_2}(\mathbf{E}_u, \mathbf{E}_v) \\ \hat{y}_{uv} &= f^3_{\Theta_3}(\mathbf{Z}_u, \mathbf{Z}_v) \end{aligned} \tag{2}$$

where $f^1_{\Theta_1}$ represents the KRL model with parameter $\Theta_1$ to convert the waveform knowledge into a low-dimensional embedding representation vector. $f^2_{\Theta_2}$ represents the ERE model with parameter $\Theta_2$ to enhance the features of the above vector. $f^3_{\Theta_3}$ represents the CF model with parameter $\Theta_3$ to calculate the probability score $\hat{y}_{uv}$. Finally, the waveform candidates are sorted and recommended based on the probability scores.

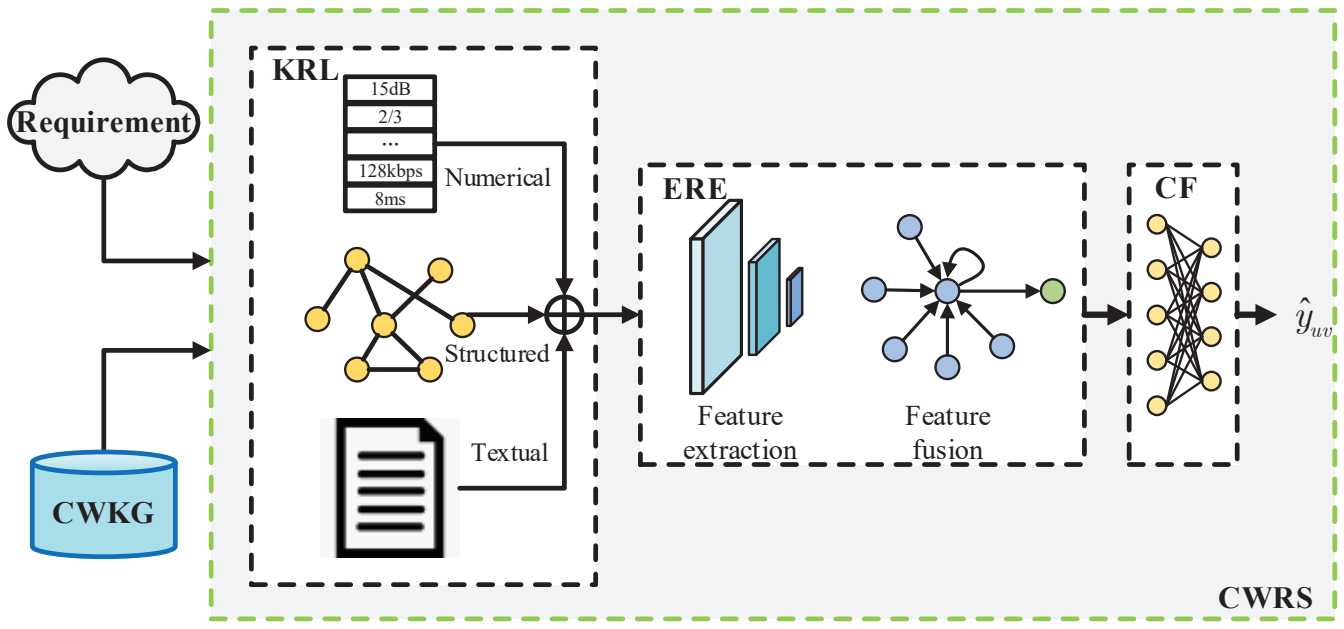

Fig. 2. The framework of the CWRS

## III. Knowledge Graph Representation of Communication Waveform

Generally, the communication waveform knowledge includes both the waveform scheme and its performance under a specific environment. A communication waveform scheme consists of many key components with corresponding parameters, e.g., modulation and coding, multi-carrier or single carrier, spread spectrum or not, etc. Meanwhile, the performance of a communication waveform is associated with the application scenarios, which are usually described by the propagation channels. On the other hand, some unintended interference or malicious jamming may also exist in the environment. The information transmission requirement of a communication system includes data rate, bit error rate (BER), demodulation signal-to-noise ratio (SNR), etc.

To uniformly and comprehensively describe the communication waveform knowledge, we utilize the knowledge graph to construct a CWKG. The CWKG consists of three parts, including the Waveform Knowledge Graph (WKG), Environment Knowledge Graph (EKG), and Environment-Waveform Bipartite Graph (EWBG).

### *A. Waveform Knowledge Graph (WKG)*

In the WKG, a waveform is represented by the head entity $h_v$, various parameter types of the waveforms are defined as relationship $r_v$, and the specific values of waveform parameters are defined as tail entity $t_v$. All the head and tail entities of the waveforms constitute the waveform entity set $\mathcal{E}_v$, and all relationships constitute the relationship set $\mathcal{R}_v$. Therefore, the WKG can be expressed as triplets $\{(h_v, r_v, t_v) \,|h_v, t_v \in \mathcal{E}_v, r_v \in \mathcal{R}_v\}$.

### B. Environment Knowledge Graph (EKG)

In the EKG, the head entity $h_u$ is defined as a virtual or composite ***environment*** that includes the application environment and QoS requirements. For simplicity of notation, we use '***environment***' to denote this virtual environment. Various parameter types of the ***environment*** are defined as relationship $r_u$, and the specific values of parameters are defined as tail entity $t_u$. Similar to the WKG, the EKG can be expressed as triplets $\{(h_u, r_u, t_u) \,|h_u, t_u \in \mathcal{E}_u, r_u \in \mathcal{R}_u\}$.

### C. Environment-Waveform Bipartite Graph (EWBG)

To connect the WKG and EKG, we further construct an EWBG, in which the virtual ***environment*** node and the waveform node are connected through a "feasible" relationship to form a triplet $(h_u, \text{feasible}, h_v)$ if the waveform can meet the information transmission QoS requirements of the communication system in a certain environment. Therefore, the CWKG can be expressed as $\mathcal{G} = \{(h, r, t) \,|h, t \in \mathcal{E}, r \in \mathcal{R}\}$, where $\mathcal{E} = \mathcal{E}_u \cup \mathcal{E}_v$, $\mathcal{R} = \mathcal{R}_u \cup \mathcal{R}_v \cup \{\text{feasible}\}$. For consistency, unless otherwise specified, the CWKG is the combination $\mathcal{G}$ of WKG, EKG, and EWBG as shown in Fig. 3.

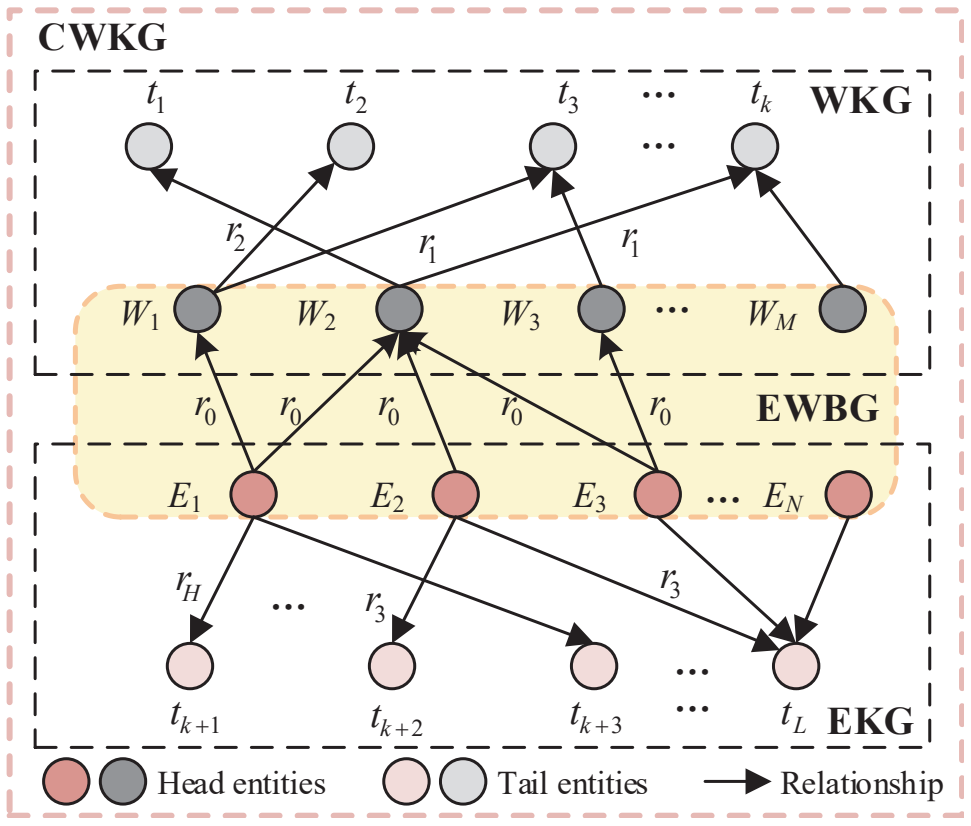

Fig. 3. Schematic diagram of the CWKG. $E_i$ and $W_i$ are head entities. $r_i$ and $t_i$ are relationships and tail entities, respectively

In Fig. 3, we use $(E_1, r_0, W_1)$ to illustrate that the waveform $W_1$ can match ***environment*** $E_1$. Obviously, $E_1$ and $W_1$ consist of various parameters. For example, $(E_1, r_H, t_{k+3})$ can denote the jamming type $r_H$ of ***environment*** $E_1$ to be multi-tone jamming $t_{k+3}$. Meanwhile, $(W_1, r_2, t_2)$ can denote the modulation $r_2$ of waveform $W_1$ to be QPSK $t_2$. Some typical relationship types involved in the waveform and the ***environment*** are shown in TABLE I.

In summary, the proposed knowledge graph representation of communication waveforms has the following features:

- By representing the parameters of waveforms and ***environments*** in triplets, the difficulty of characterizing the communication waveform knowledge in high dimensional space is avoided.
- Benefit from the same structure of WKG and EKG, a unified method can be used to perform data mining and fusion of the ***environment*** and waveform information.
- By building the CWKG with first-order neighbors, the embedded propagation can be efficiently implemented.

TABLE I
THE RELATIONSHIP TYPES IN THE CWKG

| | Relationship | Values |
|---|---|---|
| Waveform | CRC | CRC-4, CRC-8, ... |
| | Modulation | BPSK, QPSK, MSK, ... |
| | Coding type | RS, Turbo, LDPC, ... |
| | Coding rate | 2/3, 1/2, 1/3, ... |
| | ... | ... |
| | Jamming suppression | on/off |
| | Soft demodulation | on/off |
| | Bit rate(Rb) | 128kbps, 2Mbps, 64Mbps, ... |
| ***environment*** | Channel type | Gaussian, Rician, Rayleigh, ... |
| | Jamming type | single-tone, multi-tone, partial-band,Gaussian pulse, ... |
| | Num of tones | single-tone: 1, multi-tone: >1 |
| | Bandwidth factor | 0∼1 |
| | ... | ... |
| | JSR(dB) | 0, 5, 10, ... |
| | $E_b/N_0$(dB) | 0, 1, 2, 3, ... |
| | Bit rate(Rb) | 128kbps, 2Mbps, 64Mbps, ... |

## IV. Waveform Candidates Generation Based on the Proposed CWRS

As described in the previous section, the triplets in the CWKG represent the scheme with associated parameters of the existing communication waveforms and their achievable performance in a specific environment. By utilizing the developed CWKG, we describe the details of the CWRS to generate waveform candidates for a new ***environment*** in this section. As represented in the CWKG, the new ***environment*** consists of the propagation environment and QoS requirements.

### A. Knowledge Representation Learning

As shown in Fig. 2, KRL is the first step and critical part of the CWRS, which is responsible for mapping the triplets in CWKG into a low-dimensional embedding representation vector. In this paper, the widely used TransD model [8] is adopted. However, as the TransD model is responsible for the structured information, the Word2vec model is adopted to handle the textual information, e.g., the word 'Turbo's plain meaning. In contrast to the knowledge graph used in other fields, numerical information, such as coding rate, is crucial for communication waveforms. Therefore, we simultaneously fuse textual information, numerical information, and structured information to obtain a multi-channel low-dimensional embedding representation vector shown in Fig. 4.

Specifically, for the waveform, the numerical feature vector, the output vectors of the TransD model, and the Word2vec model are cascaded into a 3-channel embedding representation vector $\mathbf{E}_v$ with dimension $N_v \times N_{emb} \times 3$. Here, $N_v$ is the number of a waveform's feature, and $N_{emb}$ is the dimension of embedding representation. For the ***environment***, an embedding representation vector $\mathbf{E}_u$ with dimension $N_u \times N_{emb} \times 2$ is obtained, where $N_u$ is the number of ***environment***'s features.

Note that the input of the ***environment*** may introduce new entities when applying the trained network to recommend waveform candidates. To avoid re-training the TransD model, the corresponding embedding representation vectors are not included for the ***environment***. For example, the ***environment*** changes when a new jamming type arises, and then we have to re-train the TransD model to learn the representation of the new entity. However, these issues barely occur in the waveform scheme because it remains quasi-static for a long time.

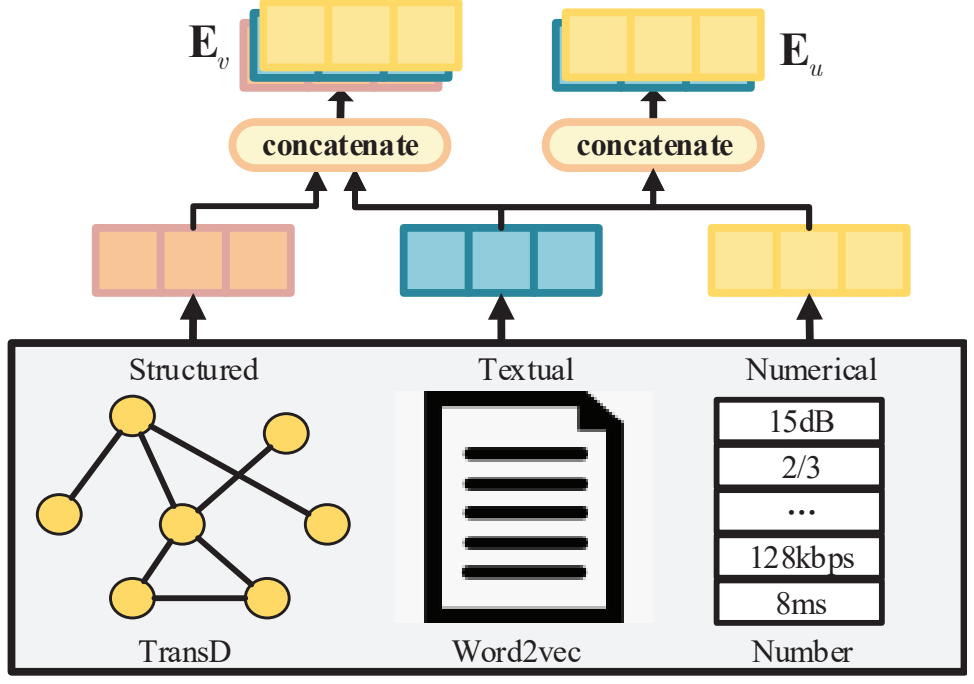

Fig. 4. Multi-channel representation of communication waveform knowledge. Specifically, '15dB' means the JSR is 15dB, '2/3' means the coding rate is 2/3, '8ms' means the pulse period is 8ms, etc.

The initial input of KRL is the triplet $(h, r, t) \in \mathcal{G}$. The TransD model first initializes each triplet into two sets of vectors $(\mathbf{h}, \mathbf{r}, \mathbf{t})$ and $(\mathbf{h}_p, \mathbf{r}_p, \mathbf{t}_p)$, where the former is the embedding representation vector, and the latter is the projection vector. Then, the score is calculated as follows [8]:

$$\begin{aligned} \hat{f} &= \|\mathbf{h}_\perp + \mathbf{r} - \mathbf{t}_\perp\|^2 \\ \mathbf{h}_\perp &= \left(\mathbf{r}_p \mathbf{h}_p^T + \mathbf{I}\right) \mathbf{h} \\ \mathbf{t}_\perp &= \left(\mathbf{r}_p \mathbf{t}_p^T + \mathbf{I}\right) \mathbf{t} \end{aligned} \tag{3}$$

where $\hat{f}$ is the score, $\mathbf{I}$ is the identity matrix, and $\mathbf{h}_\perp$ and $\mathbf{t}_\perp$ are projected vectors of head and tail entities, respectively.

Since the triplets in the CWKG are all positive samples, the system needs to generate negative samples automatically. Specifically, two negative sample triplets are generated from each positive sample triplet by randomly replacing the head or tail entity. To make the difference of the scores between the positive sample and the negative sample as large as possible, the following Bayesian Personalized Ranking (BPR) loss function [10] is used to train the TransD model.

$$L_1 = -\frac{1}{N_s} \sum_{n=1}^{N_s} \ln \left( \sigma \left( \hat{f}_n^{pos} - \hat{f}_n^{neg} \right) \right) \tag{4}$$

where $N_s$ is the number of samples, $\hat{f}_n^{pos}$ and $\hat{f}_n^{neg}$ are the scores of the positive sample and the negative sample, respectively, and $\sigma(\cdot)$ is the Sigmoid activation function.

### B. Embedding Representation Enhancement

Intuitively, the output vectors of KRL are raw data similar to an original image. Therefore, ERE is applied to highlight the features of a waveform or ***environment***. To simultaneously consider the semantic information and keep the connection information, a propagation-based method [11] including both feature extraction and fusion is adopted for ERE.

It seems that the format of communication waveform embedding representation vector is similar to the image format in computer vision. Then, a convolutional neural network (CNN) may be considered for feature extraction in ERE. However, compared with the image, the obtained vector has the following characteristics:

- Each row of the vector is derived from independent waveform parameters [1], and there is no spatial continuity.
- Different channels of the vector describe the same object in various feature spaces.

According to the above characteristics, it is preferable to use different kernels among rows for feature extraction and share kernels among channels. Therefore, CNN is not suitable for communication waveform ERE as it is space-agnostic and channel-specific. Then, based on the involution operator presented in [12], which is space-specific and channel-agnostic, we propose a modified Involution1D operator to perform communication waveform ERE.

*1) Involution1D:* Let the previously obtained vector $\mathbf{E}_v$ or $\mathbf{E}_u$ take a slice of a coordinate point $(i, j)$ along the channel as $\mathbf{X}_{\Omega_{ij}} \in \mathbb{R}^{1\times K\times C}$, and flatten it into a vector $\mathbf{X}_{ij} \in \mathbb{R}^{1\times KC}$. The basic idea of Involution1D is to use the generating function $\varphi(\cdot)$ to generate a space-specific kernel $\mathcal{K}_{ij} \in \mathbb{R}^{1\times K\times G}$ for the input $\mathbf{X}_{ij}$, and share the generated kernel $\mathcal{K}_{ij}$ on all channels. Here, $K$ is the size of the kernel, and $G$ is the number of groups of the kernel (i.e. each spatial position shares $G$ groups of different kernels). In particular, $\varphi(\cdot)$ consists of two fully connected layers with LeakyReLU activation functions. The resulting $e_{ij}^{\text{inv}}$ for the generated kernel $\mathcal{K}_{ij}$ and the corresponding input $\mathbf{X}_{\Omega_{ij}}$ is given as follows,

$$e_{ij}^{\text{inv}} = \sum_{g=1}^{G} \sum_{k=1}^{K} \left( \mathcal{K}_{ij} [\times] \mathbf{X}_{\Omega_{ij}} \right) \tag{5}$$

where the operator '$[\times]$' represents the matrix expansion product. It is defined as follows:

$$(\mathbf{A}[\times]\mathbf{B})_{kgc} = a_{kg} b_{kc} \tag{6}$$

where $a_{kg}$ is the $(k, g)$th element of matrix $\mathbf{A} \in \mathbb{R}^{K\times G}$, and $b_{kc}$ is the $(k, c)$th element of matrix $\mathbf{B} \in \mathbb{R}^{K\times C}$.

*2) Multi-head Self-Attention:* Because various waveform parameters have different influences on the communication waveform, it is necessary to emphasize the crucial waveform features by fusing various waveform parameters with different weights. It is difficult for graph convolution networks to assign different weights to various neighbor nodes, so we adopt an attention mechanism to merge various waveform parameters. Since the source and target of our problem are the same, the self-attention mechanism [13] is adopted. For a given input $\mathbf{X}$, the self-attention is performed as follows:

$$\mathbf{Z}_{\text{att}} = \text{softmax}\left(\mathbf{Q}\mathbf{K}^{\text{T}}\right)\mathbf{V} \tag{7}$$

where $\mathbf{Q}$, $\mathbf{K}$, $\mathbf{V}$ are vectors of query, key, and value, respectively. All of them are generated by the linear transformation of $\mathbf{X}$, and $\mathbf{X}$ is the output $\mathbf{E}_v^{\text{inv}}$ or $\mathbf{E}_u^{\text{inv}}$ of Involution1D.

Furthermore, we use a multi-head mechanism [2] to improve the performance [13]. Each self-attention is treated as a head, and different heads can project the input to various represen-tation subspaces. As a specific subspace is more prominent for some waveform parameters than others, combining multiple subspaces can highlight the waveform parameters with significant influence. The final output is given as follows:

$$\mathbf{Z} = \mathbf{Z}_1 \oplus \mathbf{Z}_2 \oplus \cdots \oplus \mathbf{Z}_H \tag{8}$$

where $\mathbf{Z}_i$ represents the output of the $i$th self-attention, $H$ is the number of heads, and '$\oplus$' represents a cascading operation.

[1]Note that this is also true for both environment and requirement parameters. Only waveform parameters are discussed below for brevity.

[2]The multi-head mechanism has been successfully applied in the field of natural language processing (NPL).

### C. Collaborative Filtering

As mentioned before, we apply KRL and ERE to characterize the ***environment*** and the waveform, respectively. The characteristics of the waveform include both its label and waveform parameters. To fully evaluate the matching degree between the ***environment*** and the waveform, the MLP-based CF [14] is adopted.

For the MLP-based CF, the characteristic representations $\mathbf{Z}_u$ and $\mathbf{Z}_v$ of the ***environment*** and the waveform are first cascaded. Then, the matching degree $s_{uv}$ is learned by feeding the resulting vector into the MLP. That is,

$$\begin{aligned}
\mathbf{Z}_0^{\mathrm{MLP}} &= \mathbf{Z}_u \oplus \mathbf{Z}_v \\
\mathbf{Z}_1^{\mathrm{MLP}} &= a_1\left(\mathbf{Z}_0^{\mathrm{MLP}}\mathbf{W}_1 + \mathbf{b}_1\right) \\
&\cdots \\
\mathbf{Z}_L^{\mathrm{MLP}} &= a_L\left(\mathbf{Z}_{L-1}^{\mathrm{MLP}}\mathbf{W}_L + \mathbf{b}_L\right) \\
s_{uv} &= \mathbf{h}^T \mathbf{Z}_L^{\mathrm{MLP}}
\end{aligned} \tag{9}$$

where $\mathbf{W}_i$, $\mathbf{b}_i$, $a_i$ are weight matrix, bias vector, and activation function of the $i$th perceptron, respectively. $\mathbf{h}$ is a learnable coefficients vector. Finally, the probability of each waveform concerning the ***environment*** is calculated as follows:

$$\hat{\mathbf{y}}_u = \mathrm{softmax}\left(\mathbf{s}_u\right) \tag{10}$$

where $\mathbf{s}_u = \left[s_{u1}, s_{u2}, ..., s_{uM}\right]^T$, $M$ is the total number of waveforms.

### D. Network Training Process

For an ***environment***, the score for an available waveform is 1, or 0 for unavailable or unknown waveforms. Then, the score $y_{uv} \in \{0, 1\}$, which is similar to the binary classification task. Therefore, we adopt the cross-entropy loss function in network training. It is given as follows:

$$\begin{aligned}
L_2 &= -\sum_{(u,i)\in\mathbf{O}^+} \ln\left(\hat{y}_{ui}\right) - \sum_{(u,j)\in\mathbf{O}^-} \ln\left(1 - \hat{y}_{uj}\right) \\
&= -\sum_{(u,i)\in\mathbf{O}} \left(y_{ui}\ln\left(\hat{y}_{ui}\right) + \left(1 - y_{ui}\right)\ln\left(1 - \hat{y}_{ui}\right)\right)
\end{aligned} \tag{11}$$

where $\mathbf{O} = \mathbf{O}^+ \cup \mathbf{O}^-$, $\mathbf{O}^+$ represents the sample set of available waveforms, $\mathbf{O}^-$ is the sample set of unavailable or unknown waveforms, and satisfies $|\mathbf{O}^+| = |\mathbf{O}^-|$. We use the Adam optimizer to alternately optimize the loss $L_1$ in (4) and $L_2$ in (11).

## V. Simulation Results

Up to now, we have built a CWKG consisting of 46 waveform entities and 12470 ***environment*** entities. The waveforms can be divided into two categories, namely Link16 data link single-carrier waveforms [15] and OFDM waveforms. The jamming involved in the ***environment*** includes single-tone, multi-tone, partial-band, and Gaussian pulse jamming [16].

In the simulations, the hyper-parameters are set as follows: the learning rate of the Adam optimizer is $lr = 0.001$, and the parameters of Involution1D are $K = 5, G = 1$. The training set and test set are allocated at a ratio of 10:2, and the recommendation success rate is measured by $Hit@1$. It is defined as:

$$Hit@1 = \frac{N_{Hit}}{N_{All}} \tag{12}$$

where $N_{Hit}$ is the number of successful recommendations, and $N_{All}$ is the total number of recommendations. Specifically, if the recommended waveform with the highest probability score belongs to the set of available waveforms, it will be counted as a successful recommendation. For example, when the ***environment*** consists of additive white Gaussian noise (AWGN) channel, single-tone jamming with jamming-to-signal power ratio (JSR) 30dB, the required date rate $R_b$=5Mbps with bit-error-rate (BER) $10^{-6}$ and demodulation threshold $E_b/N_0$=4dB, our CWRS recommends an OFDM waveform candidate supporting $R_b$=6.8246Mbps and consisting of CRC-64, QPSK, 1/3 Turbo coding with a coding length 4160, random interleaver with interleave length 175616, soft demodulation and frequency-domain jamming suppression. The average success rate presented in TABLE II-IV is computed based on the last 100 epochs after the system converges.

First, we investigate the effect of ERE with different methods shown in Fig. 5. Compared with the performance without ERE marked as 'KRL only', it can be found that the recommendation success rate can be significantly improved by introducing ERE. The converged average success rates are presented in TABLE II. It can be observed that the ERE with self-attention and involution1D can achieve nearly the same performance. Meanwhile, the performance will significantly deteriorate if the ERE with CNN is used. As discussed in Section IV, the Involution1D is more preferable compared to CNN for our proposed CWKG.

Subsequently, we investigate the effect of multi-head self-attention. Fig. 6 and TABLE III present the performance comparison for different head numbers. When self-attention is only used in ERE, it can be seen that the performance can be improved with the increase of $H$ at the cost of computational complexity. As the performance gain is negligible when the number of heads is greater than 3, $H$=3 is a reasonable choice.

Finally, we present the performance of the whole system in Fig. 7 and TABLE IV. We consider several cascading schemes of involution1D and multi-head self-attention for ERE. In Fig. 7 and TABLE IV, 'Invo1D+($H$=1)' means that Involution1D is first used to extract features, and then multi-head self-attention with $H$=1 is used for feature fusion. It can be found that better performance can be achieved if involution1D is first used to extract the features. Compared with the performance given in TABLE III, the cascading scheme with 'Invo1D+($H$=3)' can achieve the performance close to the multi-head self-attention with $H$=8. Therefore, by cascading involution1D and multi-head self-attention, the multi-head number can be significantly reduced to achieve nearly the same performance.

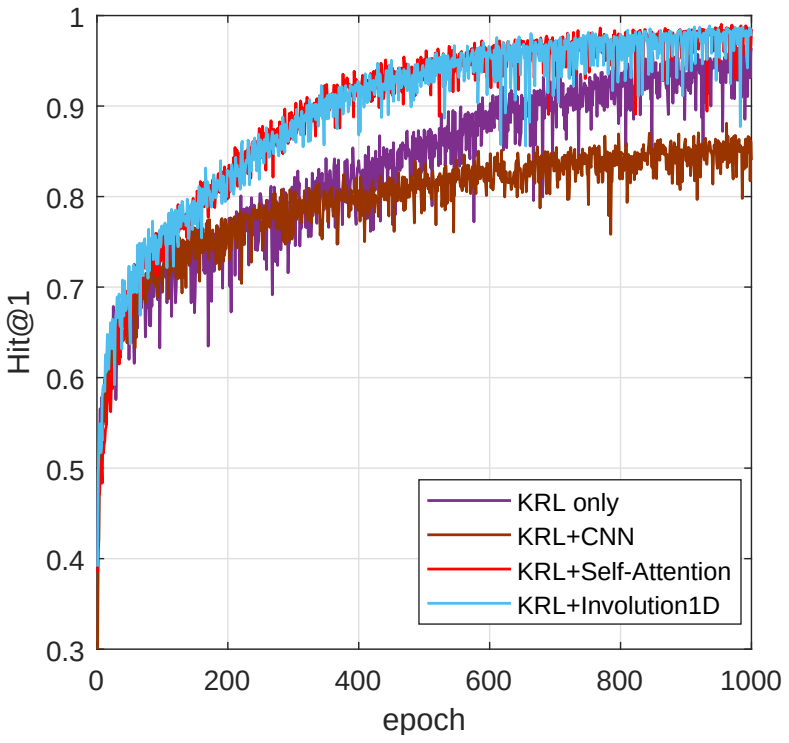

Fig. 5. The effect of ERE methods

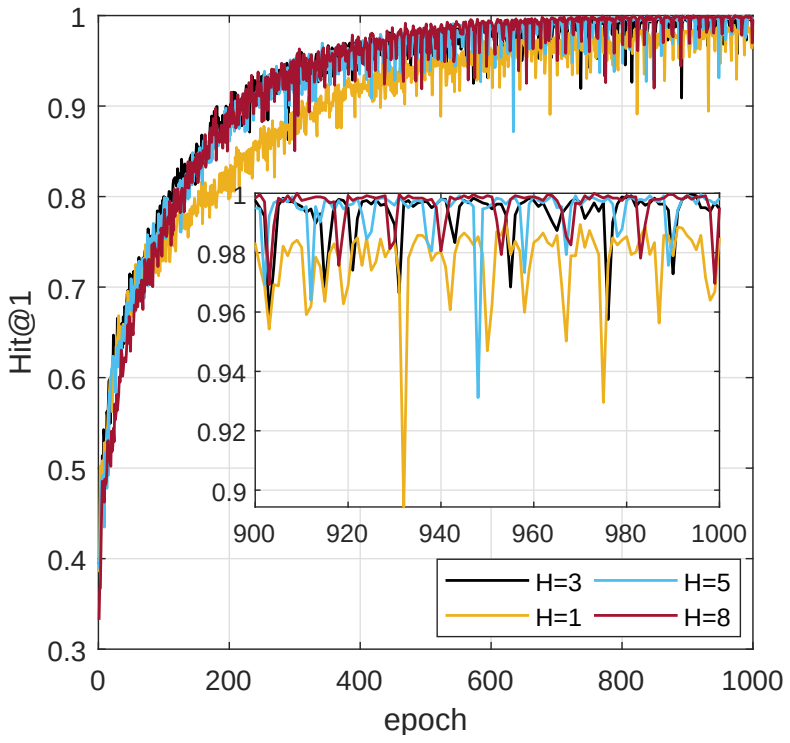

Fig. 6. The effect of multi-head numbers

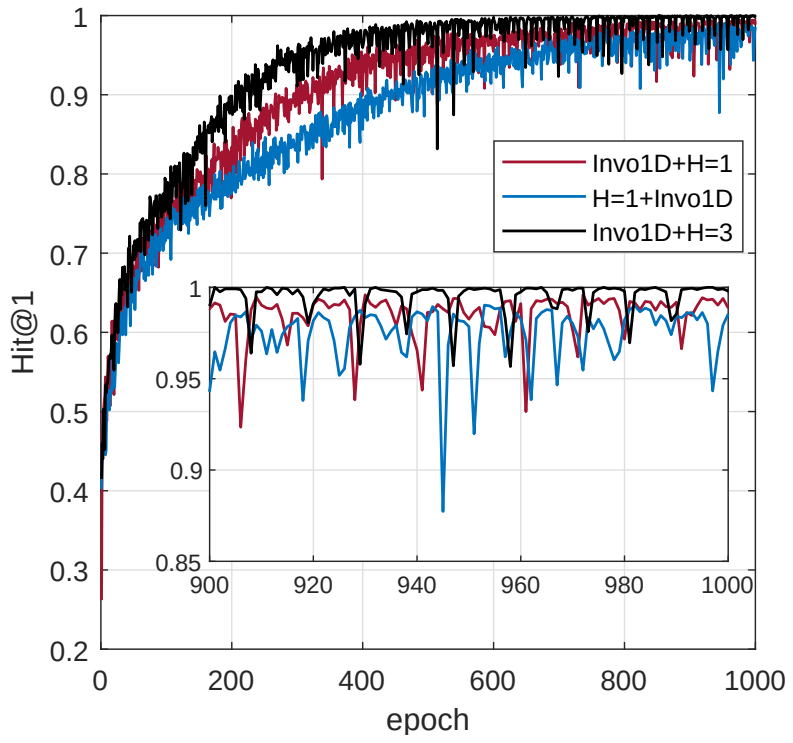

Fig. 7. The effect of cascading schemes

TABLE II
THE PERFORMANCE OF DIFFERENT ERE METHODS

| **Method** | KRL only | KRL+CNN | KRL+ Self-Attention | KRL+ Invotion1D |
|---|---|---|---|---|
| $Hit@1$ | 0.9401 | 0.8479 | 0.9771 | 0.9741 |

TABLE III
THE PERFORMANCE OF DIFFERENT MULTI-HEAD NUMBERS

| $H$ | 1 | 3 | 5 | 8 |
|---|---|---|---|---|
| $Hit@1$ | 0.9771 | 0.9933 | 0.9943 | 0.9959 |

TABLE IV
THE PERFORMANCE OF DIFFERENT CASCADING SCHEMES

| **Mode** | Invo1D+($H$=1) | ($H$=1)+Invo1D | Invo1D+($H$=3) |
|---|---|---|---|
| $Hit@1$ | 0.9853 | 0.9752 | 0.9951 |

## VI. CONCLUSION AND FUTURE RESEARCH

In this paper, we propose a new communication waveform design paradigm based on a communication waveform knowledge graph (CWKG). By considering both waveform schemes and their performance under specific application environments, we build a CWKG integrating structured, textual, and numerical information of communication waveform knowledge. Based on the developed CWKG, we further propose a communication waveform recommendation system (CWRS) to generate waveform candidates for new requirements specified by both environment and QoS. By taking into consideration the characteristics of CWKG, our proposed CWRS consists of three parts: knowledge representation learning (KRL), embedding representation enhancement (ERE), and multilayer perceptron (MLP)-based collaborative filtering (CF). Simulations show that the proposed CWKG based CWRS can achieve a considerably high recommendation success rate.

As mentioned in Section II, this paper mainly focuses on the waveform recommendation with the CWKG of the proposed paradigm. In future research, we will further expand the CWKG and add waveform optimization to the system for coping with the case that the recommended waveform could not fulfill the new requirement.